\documentclass[aps,prl,amssymb,showpacs,twocolumn,floatfix]{revtex4}
\usepackage{graphicx}
\newcommand{\rsun}{R_\odot}
\begin{document}
\title{Shining light through the Sun}
\author{Malcolm Fairbairn$^{1,2}$,
Timur Rashba$^{3,4}$
and
Sergey Troitsky$^5$}
\affiliation{$^1$ Cosmology, Astroparticle Physics and String Theory,
Stockholm University, Sweden}
\affiliation{$^2$ Perimeter Institute,
Waterloo, Ontario, Canada N2L 2Y5 }
\affiliation{$^3$ Max-Planck-Institut f\"ur Physik
(Werner-Heisenberg-Institut), F\"ohringer Ring 6, D-80805 M\"unchen,
Germany}
\affiliation{$^4$ Pushkov Institute
(IZMIRAN), Troitsk 142190, Russia}
\affiliation{$^5$ Institute for Nuclear Research of RAS, 60th October
Anniversary Prospect 7a, Moscow 117312, Russia}
\pacs{14.80.Mz,98.70.Rz,95.10.Gi}

\begin{center}
\begin{abstract}
It is shown that the Sun can become partially transparent to high energy
photons in the presence of a pseudo-scalar.  In particular, if the axion
interpretation of the PVLAS result were true then up to 2\% of GeV energy
gamma rays might pass through the Sun, while an even stronger effect is
 expected for some axion parameters. We discuss the possibilities of
 observing this effect. Present data are limited to the observation of
the solar occultation of 3C~279 by EGRET in 1991; 98\% C.L.\ detection of
a non-zero flux of gamma rays passing through the Sun is not
yet conclusive. Since the same occultation happens every October, future experiments, e.g.\ GLAST, are expected to have
better sensitivity.
\end{abstract}
\end{center}
\maketitle
Many contemporary models of particle physics predict the existence of
light pseudoscalar bosons which are often refereed to as axion-like
particles, or simply axions.

The Lagrangian density of the photon-pseudoscalar system is given by
\begin{equation}
\mathcal{L}=\frac{1}{2}(\partial^\mu a\partial_\mu a-m^2a^2)
-\frac{1}{4}\frac{a}{M}F_{\mu\nu}\widetilde
F^{\mu\nu}-\frac{1}{4}F_{\mu\nu}F^{\mu\nu}
\label{lag}
\end{equation}
where $F_{\mu\nu}$ is the electromagnetic stress tensor and $\widetilde
F_{\mu\nu}=\epsilon _{\mu \nu \rho \lambda }F_{\rho \lambda }$ its dual,
$a$ the pseudoscalar (axion) field, $m$ the axion mass and $M$ is its
inverse coupling to the photon field. The coupling of the photon and
pseudoscalar fields in this way means that a photon has a finite
probability of mixing with its opposite polarisation and with the
pseudoscalar in the presence of an external magnetic field
\cite{sikivie,raffelt}.
One way of searching for such interactions is to pass photons through a
magnetic field within which some conversion into pseudoscalars would be
expected given a Lagrangian such as (\ref{lag}).  The beam is then
directed through an opaque medium, a wall for instance, and then passed
though another magnetic field where any pseudoscalars which were created
may convert back into photons.
Experiments which aim to 'shine light through walls' in this way are more
likely to succeed if one uses higher energy photons and stronger magnetic
fields.   In this letter we propose using the Sun as our magnetic field
and as our wall, and gamma-ray sources from outside the solar system for
our photons. This is feasible because the Sun does not appear to emit
gamma-rays itself (at least in its quiet phase)~\cite{EGRET:SunMoon}.

Recently, the interest in axion-like particles has been re-ignited by the
results of the PVLAS experiment which has detected a rotation in the plane
of polarization of a laser passing through a magnetic field \cite{pvlas}.
The level and nature of the effect is different from expected quantum
electro-dynamics effects.
It is claimed that the PVLAS result
is compatible with the existence of such a pseudoscalar with a mass of
$m\sim 10^{-3}$ eV and an inverse coupling $M\sim 10^5$ GeV.  This is
unexpected since the CAST~\cite{CAST} experiment has ruled out such a strongly coupled axion.  
Some authors have therefore suggested that there may be a composite pseudoscalar \cite{composite} with a form factor which suppresses its thermal production in the core of the sun while leaving its classical mixing properties unchanged.  Since CAST looks for thermally produced axions, the compositeness is seen as a way of reconciling the results from CAST and PVLAS (see also\cite{cooked}). 

It is feasible that in a very few years, the axion interpretation of the PVLAS
result may be tested using x-ray lasers and de-commissioned magnetic fields
from particle accelerators \cite{ringwald}.  Here we suggest that it might
be tested on a similar time scale by searching for gamma rays from
astrophysical objects when they are eclipsed by the Sun. (see also \cite{zioutas,dupays} for related work)

We are interested in spatial variations in the photon state vectors so we
expand the photon field $A(t,x)$ in components of fixed frequency
$A(x)e^{-i\omega t}$. If the magnetic field changes on length scales much
larger than the wavelength of the particles and the refractive index
$|n-1|\ll 1$ then one can expand~\cite{raffelt} the operators in the
equation of motion
for the three species $\omega^2+\partial_z^2\rightarrow
2\omega(\omega-i\partial_z)$  and write down the
Schr\"{o}dinger equation
\vspace{-1mm}
\begin{equation}
i\partial_z \Psi=
-\left(\omega+\mathcal{M}\right)\Psi\qquad;\qquad
\Psi=\left(A_{x},A_{y},a\right)
\label{schrodinger}
\end{equation}
where the matrix $\mathcal{M}$ is given by
\begin{equation}  \label{mixmat}
{\cal M}\equiv\left(
\begin{array}{ccccccccc}
\Delta_p&0&\Delta_{Mx}\\
0&\Delta_p&\Delta_{My}\\
\Delta_{Mx}&\Delta_{My}&\Delta_m
\end{array}
\right)\hspace{0.7cm}
\end{equation}
the mixing parameters are the refractive index due to
electrons in medium $\Delta_p$, the mass term for the pseudoscalar
$\Delta_m$ and the off-diagonal which give the strength of the mixing
$\Delta_M$.  The three parameters take values
$$
\begin{array}{rcccl}
\displaystyle
\Delta_{M\! i}
\!\!\!&=&\!\!\!
\displaystyle
\frac{B_i}{2M}
\!\!\!&=&\!\!\!
\displaystyle
1.755\times 10^{-11}\left(\frac{B_i}{1~{\rm G}}\right)
\left(\frac{10^{5}~{\rm GeV}}{M}\right){\rm cm}^{-1}\!,\\
\displaystyle
\Delta_m
\!\!\!&=&\!\!\!
\displaystyle
\frac{m^2}{2 \omega}
\!\!\!&=&\!\!\!
\displaystyle
2.534\times 10^{-11}\left(\frac{m}{10^{-3}~{\rm eV}}\right)^2
\left(\frac{1~{\rm GeV}}{\omega}\right){\rm cm}^{-1}\!,\\
\displaystyle
\Delta_p
\!\!\!&=&\!\!\!
\displaystyle
\frac{\omega _p^2}{2 \omega }
\!\!\!&=&\!\!\!
\displaystyle
3.494\times
10^{-11}\!\!\left(\frac{n_e}{10^{15}~\rm{cm}^{-3}}\right)
\!\left(\frac{1~{\rm GeV}}{\omega}\right){\rm cm}^{-1}\!,
\end{array}
$$
where $\omega _p=4\pi \alpha n_e/m_e$ is the plasma frequency,
$n_e$ is the electron density,
$B$ is the magnetic field, $m_e$ is the electron mass, $\alpha $ is
the fine-structure constant, $\omega $ is the photon's (axion's) energy.
For constant magnetic field and electron density, the axion-photon
conversion probability is given by
\begin{equation}
P=\frac{4 B^2 \omega^2}{M^2\left(\omega _p^2-m^2 \right)^2+4 B^2
\omega^2} \sin^2\left(
\pi \frac{z}{l_{\rm osc}}
\right),
\label{P}
\end{equation}
where the oscillation length is
$$
l_{\rm osc}=
\frac{4\pi \omega M}{\sqrt{M^2
\left(\omega _p^2-m^2 \right)^2
+4 B^2 \omega^2}},
$$
and
$z$ is the coordinate along the path. The resonant conversion takes place
at $\omega _p\sim m$, when the conversion probability is of order one
provided the size of the resonant region
$\Delta z_{\rm res}$
is much larger than
the oscillation length $l_{\rm osc}$.

\begin{figure}[t!]
\begin{center}
\includegraphics[width=0.45\textwidth,angle=0]{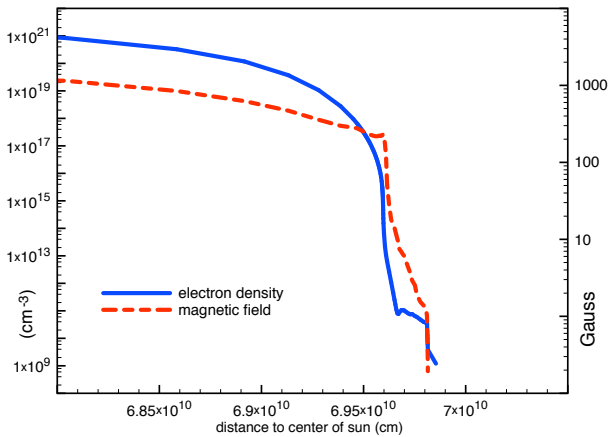}
\caption{\it Electron density and magnetic field near the surface of the Sun and into the
convection zone.
}
\label{elec}
\end{center}
\end{figure}

The solar model we choose is a combination of two models, one
by Spruit~\cite{spruit} for the convection zone (we go no deeper here) and one by Vernazza
\cite{vernazza} for the chromosphere and photosphere.  The electron
density is shown in figure \ref{elec}.
The magnetic fields inside the Sun are poorly known.
We assume that they are in equipartition with the bulk motion
of the plasma rather than the electron pressure, which gives us a clearly
defined average value of the magnetic field at each radius within the Sun.
However,
the magnetic field inside the Sun is turbulent, and it is necessary
to include this variation in the magnetic field along the line of flight
of the photon/pseudoscalar to obtain realistic mixing.
We therefore
assume random magnetic fields in the $x$ and $y$ direction generated
by the Fourier transform of a power spectrum with random phases. For this
power spectrum, we choose a flat spectrum with a short distance cut-off at
100 km. The variance of the magnetic field is normalised to be equal to
the magnetic field at each radius.
We see that the electron density changes by 8 orders of
magnitude in a layer of some 2000~km below the solar surface, thus
providing conditions for the resonance for a wide region of the axion
parameter space. On the other hand, this region is close enough to the
surface so one may expect that the probability for a photon, created
in the resonant region, to escape from the Sun, is significant.

For the
energy range we consider, namely gamma rays with $E>100$~MeV, the
interaction between the photon and the medium is dominated by the pair
production on the nuclear electric field well described by the
Bethe-Heitler cross section.

\begin{figure}
\begin{center}
\includegraphics[width=0.45\textwidth,angle=0]{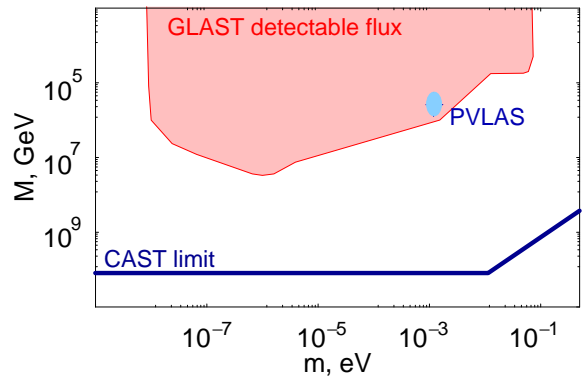}
\caption{\it
Axion parameter space. For the shaded region, GeV photon
flux through the Sun detectable with GLAST at the solar occultations of 3C~279 is expected.
The ellipse corresponds to PVLAS-motivated parameters~\cite{pvlas}.
The values of $M$ above the thick line are disfavored by
CAST~\cite{CAST}.}
\label{fig:parameters}
\end{center}
\end{figure}

To obtain indications of the regions in parameter space where the maximal
mixing
is allowed, we estimate the location and size of the resonant zone in the
adiabatic approximation with the help of Eq.~(\ref{P}) and then impose two
requirements: that the resonance takes place in the optically thin zone and
that $\Delta z_{\rm res}>l_{\rm osc}$. Quite straightforwardly, one obtains
that the maximal mixing is allowed for relatively strongly coupled axions
of various masses (see Fig.~\ref{fig:parameters}).

Inside the Sun, away from the resonance, the conversion probability is
very small. Despite this, the high density and almost immediate
absorption of any occasionally created photon may result in cumulative
decrease of the axion flux. In the same adiabatic approximation, for the
gamma-ray energies $E \lesssim 10$~GeV and
for the paths not going very deep into the Sun (impact
parameters $d\gtrsim 0.7 \rsun$), this cumulative effect does not reduce
the number of axions in the beam by more than 10\%. This means that for
the parameters specified, a non-negligible part of the initial
photon-axion mixture would travel through the optically thick regions in
the form of axions and would convert back to photons at the solar surface.
It is therefore possible that a gamma ray source would be able to shine
through the Sun, although in practice only a fraction of photons would
emerge.

\begin{figure}[t!]
\begin{center}
\includegraphics[height=7.5cm,width=0.45\textwidth,angle=0]{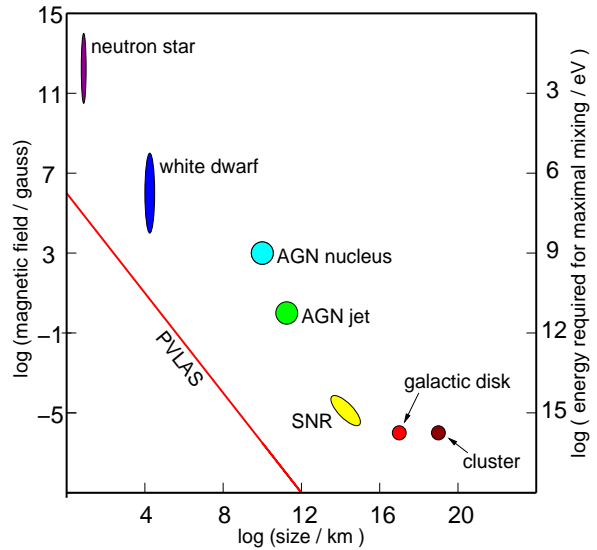}
\caption{\it Hillas plot of the magnetic field of objects vs. their size.
The oscillation length at maximal mixing for the PVLAS coupling $M=10^5$
GeV is plotted as a function of magnetic field (diagonal line).  The
right hand axis shows the photon energy required for maximal mixing as a
function of magnetic field assuming the PVLAS coupling and mass
$m=10^{-3}$ eV.}
\label{hillas}
\end{center}
\end{figure}

To be more precise and to go beyond the adiabatic approximation, we
integrate numerically the equations of motion
along a path of several thousand kilometres through the Sun. Instead of
explicitly solving the Schr\"{o}dinger equation, it is better
to go to the interaction representation and separate out the mixing that
we are interested in from the normal propagating oscillation.  This can be
done by defining a density matrix $\rho=\Psi^* \Psi$ with an evolution
equation \cite{raffelt,deffayet}
\begin{equation}
i\partial_z \rho=[\mathcal{M},\rho]-i{\cal D}\rho
\end{equation}
where here we have simultaneously introduced a damping matrix $\cal{D}$
designed to take into account interactions between the photon part of the
wavefuction and the particles inside the Sun.  This damping matrix takes
the form
\begin{equation}
{\cal D}\equiv\left(
\begin{array}{ccccccccc}
\Gamma&\Gamma&\frac{1}{2}\Gamma\\
\Gamma&\Gamma&\frac{1}{2}\Gamma\\
\frac{1}{2}\Gamma&\frac{1}{2}\Gamma&0
\end{array}
\right)
\end{equation}
where the inverse decay length $\Gamma=n_{e}\sigma$ and we again use the
Bethe-Heitler cross section for the $\sigma $.
We assume that the beam entering the Sun has a ratio of 2:1 photon:axion
for the following reason.  Figure \ref{hillas} is a Hillas plot of the
magnetic field of some typical gamma-ray sources vs. their size (the
transient nature of gamma ray bursts means that one could not predict when
to point the detector at the Sun so we have not included them on this
plot.)  All of the sources are larger in size than the oscillation length
at maximal mixing ($l_{\rm osc}\sim M/B$) corresponding to their magnetic
fields and the coupling of the PVLAS photon.  A more detailed analysis
would look at the coherence length of the magnetic field rather than the
total extent of the object, but for the smaller objects with higher
magnetic fields, one would expect these two length scales to be closer
together.  The right hand axis of figure \ref{hillas} shows the energy
necessary for maximum mixing to be achieved.  The source we will be
looking at in this paper is a quasi-stellar object, in other words an
active galaxy, so we expect there to be strong mixing and many oscillation
lengths in the source, and the high energy photons created should
therefore be equal mixtures of both photon polarisations and the axion.

\begin{figure}[t!]
\begin{center}
\includegraphics[height=7.5cm,width=0.45\textwidth,angle=0]{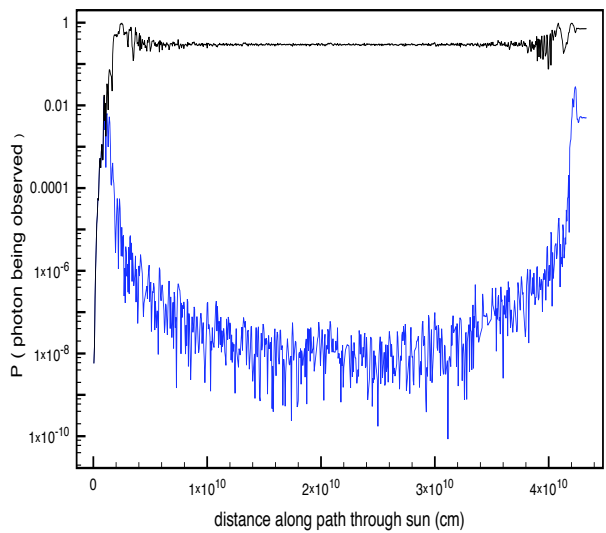}
\caption{\it  The probability of a photon emerging when a GeV PVLAS axion
enters the Sun with an impact parameter 0.95 $R_\odot$. Upper curve:
unrealistic case with photon cross section neglected.  The two resonant
regions can be clearly seen.
Lower curve: the same but
including photon interactions.}
\label{probdec}
\end{center}
\end{figure}

\begin{figure}
\begin{center}
\includegraphics[height=7.5cm,width=0.45\textwidth,angle=0]{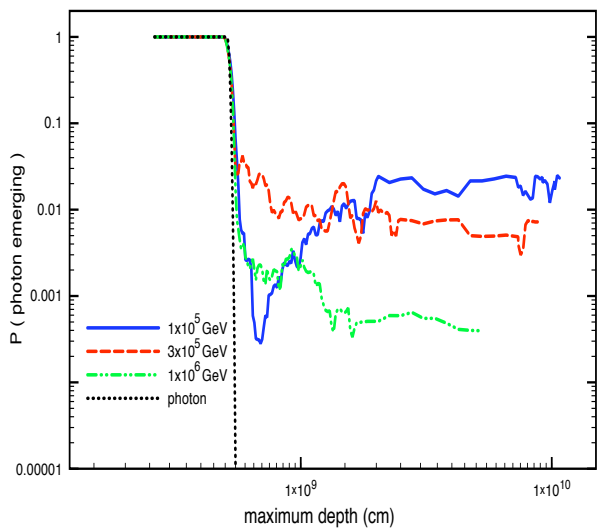}
\caption{\it The probability of a
GeV photon emerging for
pseudoscalar mass
$m=10^{-3}$ eV and different inverse coupling scales $M$.
The dotted line bounds the maximal depths still transparent for GeV
photons. }
\label{milli}
\end{center}
\end{figure}

Figure
\ref{probdec} shows the actual mixing in action along a path
through the Sun at a depth of 0.95 $R_\odot$.
In figure \ref{milli} we plot the probability of a photon emerging as a
function of the maximum depth along the path through the Sun for an axion
mass of $10^{-3}$ eV and different values of the coupling $M$.
The results of the numerical solution are in rather good agreement
with the analytical estimates presented above. In particular,
the probability of gamma rays emerging from the Sun when a bright source
passes behind it may be as high as 2\% for GeV energy photons, if the
interpretation of the PVLAS signal as an axion is valid, and even larger
for the parameters shown in Fig.~\ref{fig:parameters}.

Let us turn now to observational constraints on gamma rays passing
through the Sun. The only source in the EGRET catalogue~\cite{3EG} occulted by the Sun is the quasi-stellar object 3C~279. EGRET observed the solar occultation of 3C~279 in 1991, in the viewing period 11.0. The QSO was in moderate state
and was firmly detected in gamma rays during that viewing
period~\cite{3EG}.

The source was screened by the Sun for 8 hours 34 minutes on October 8, 1991 \cite{Planeph}.
The minimal impact parameter was $d=0.75R_\odot$.

We analysed the EGRET data following~\cite{EGRET:like}. The apparent flux of
photons with energy $E\gtrsim$100~ MeV from the location of 3C~279 during
the occultation is $(6.2^{+3.7}_{-2.7})\cdot 10^{-7}$~cm$^{-2}$~s$^{-1}$,
to be compared with the value obtained from the analysis of the rest of
the same viewing period when the source is not behind the Sun which is
$(8.6 \pm 0.5 )\cdot 10^{-7}$~cm$^{-2}$~s$^{-1}$. This latter result,
obtained from our own analysis of the data, is in a very good agreement
with the value quoted in the 3EG catalog~\cite{3EG} for this period,
$(7.94 \pm 0.75)\cdot 10^{-7}$~cm$^{-2}$~s$^{-1}$. We see that a non-zero
point-source flux from the location of 3C~279 during the occultation cannot be confirmed or
excluded at the present level of statistics. Since such a non-zero flux
could be the signal of a new elementary particle, this calls for future
astronomical observations with more precise instruments.  In particular
GLAST is predicted to have a sensitivity of $3\times
10^{-8}$cm$^{-2}$s$^{-1}$ for the period of 8 hours \cite{glast}.  By looking at the same source when it is eclipsed by the sun, GLAST will therefore be able to rule out complete transparency to gamma rays. GLAST could also observe the partial transparency predicted in this paper, confirming the interpretation of the PVLAS data as a new particle.

{\bf Acknowledgments}
Discussions:
L.~Bergstr\"om,  B.~Irby, D.~Gorbunov, V.~Rubakov, G.~Rubtsov,
M.~Tavani and P.~Tinyakov.
Funding: Vetenskapsr\aa det and PI (MF), Marie Curie program, RFBR grant 04-02-16386 and RAS Program ``Solar activity'' (TR), Russian Science Support Foundation, grants INTAS
03-51-5112 and NS-7293.2006.2, CERN (ST). Computers: Stockholm GLAST machines, INR cluster
(RF gov.contract 02.445.11.7370), PI.

\end{document}